\documentclass[showpacs,amsmath,amssymb,superscriptaddress]{revtex4}

\usepackage{amsmath}
\usepackage{amssymb}
\usepackage{amsfonts}
\usepackage{graphicx}
\usepackage{amsthm}

\newcommand{\R}{{\mathbb R}}
\newcommand{\RQ}{{\mathbb R}\setminus{\mathbb Q}}
\newcommand{\HH}{{\cal H}}

\newcommand{\ket}[1]{|{#1}\rangle}
\newcommand{\kb}[1]{|{#1}\rangle\!\langle{#1}|}

\newcommand{\bra}[1]{\langle{#1}|}

\newcommand{\psiCW}{{\varphi}}
\newcommand{\psitwoCW}{{\tilde{\varphi}}}
\newcommand{\psithreeCW}{{\varphi}^{\prime}}

\newtheorem{thm}{Theorem}

\makeatletter

\begin{document}
\title{Notes on Inhomogeneous Quantum Walks}
\author{Yutaka Shikano}
\email{shikano@th.phys.titech.ac.jp}
\affiliation{Department of Physics, Tokyo Institute of Technology, Meguro, Tokyo, 152-8551, Japan}
\affiliation{Department of Mechanical Engineering, Massachusetts Institute of Technology, Cambridge, MA 02139, USA}
\author{Hosho Katsura}
\email{hosho.katsura@gakushuin.ac.jp}
\affiliation{Department of Physics, Gakushuin University, Toshima, Tokyo 171-8588, Japan}
\affiliation{Kavli Institute for Theoretical Physics, University of California Santa Barbara, CA 93106, USA}
\date{\today}
\begin{abstract}
    We study a class of discrete-time quantum walks with inhomogeneous coins defined in [Y. Shikano and H. Katsura, 
    Phys. Rev. E {\bf 82}, 031122 (2010)]. 
    We establish symmetry properties of the spectrum of the evolution operator, which resembles the Hofstadter butterfly. 
\end{abstract}
\maketitle
Throughout this paper, we focus on
a one-dimensional discrete time quantum walk (DTQW) with two-dimensional coins.
The DTQW is defined as a quantum-mechanical analogue of the classical random walk. 
The Hilbert space of the system is a tensor product $\HH_p \otimes \HH_c$, where $\HH_p$ is the position space of a quantum walker
spanned by the complete orthonormal basis $\ket{n}$ ($n \in \mathbb{Z}$) and $\HH_c$ is the coin Hilbert space 
spanned by the two orthonormal states $\ket{L} = ( 1 , 0 )^{{\bf T}}$ and $\ket{R} = ( 0 , 1 )^{{\bf T}}$. 
Here, the superscript ${\bf T}$ denotes matrix transpose.  
A one-step dynamics is described by a unitary operator $U = WC$ with 
\begin{align}
	C &= \sum_n \left[ (a_n \ket{n,L} + c_n \ket{n,R})\bra{n,L} + (d_n \ket{n,R} + b_n \ket{n,L}) \bra{n,R} \right], 
\\
	W &= \sum_n \left( |n-1,L \rangle \langle n,L| + |n+1,R \rangle \langle n,R| \right), 
	\label{shift}
\end{align} 
where $\ket{n, \xi} =: \ket{n} \otimes \ket{\xi} \in \HH_p \otimes \HH_c \ (\xi = L, R)$
and the coefficients at each position satisfy the following relations: 
 $|a_n|^2 + |c_n|^2 = 1$, $a_n \overline{b}_n + c_n \overline{d}_n = 0$, $c_n = - \Delta_n \overline{b}_n$, 
$d_n = \Delta_n \overline{a}_n$, where $\Delta_n = a_n d_n - b_n c_n$ with $ |\Delta_n|= 1$.  
Two operators $C$ and $W$ are called coin and shift operators, respectively. 
The probability distribution at the position $n$ at the $t$th step is then defined by 
\begin{equation}
	\Pr (n;t) = \sum_{\xi \in \{ L,R \} } \left| \bra{n,\xi} U^{t} \ket{0,\phi} \right|^2.
\end{equation}
A homogeneous version of this 
DTQW was first introduced in Ref.~\cite{Ambainis}. 

\begin{figure}[tb]
	\includegraphics[width=10cm]{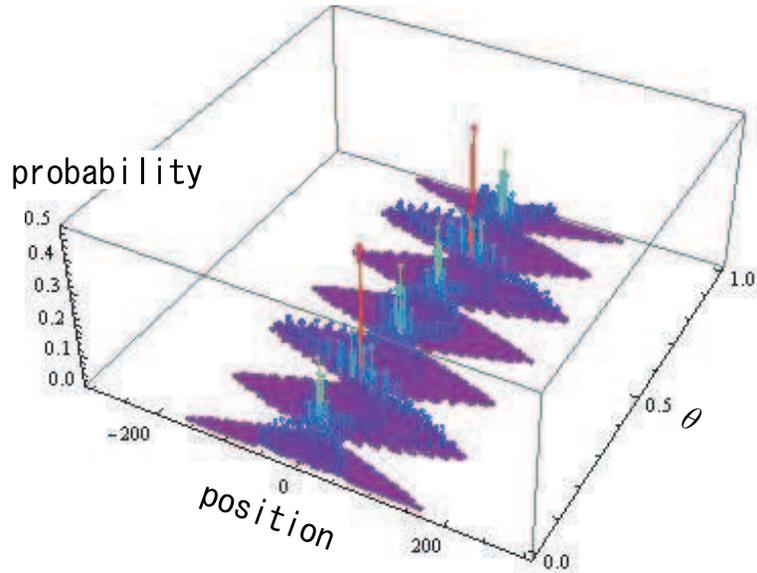}
	\caption{Probability distribution of the inhomogeneous QW at $300$th step with $\alpha=1/3$. 
    From simple algebra, it can be easily shown that
	the inhomogeneous QW is finitely confined when $\theta = (2m-1)/12 \ (m \in \mathbb{Z})$.}
	\label{open}
\end{figure}

Suppose that the coin operator is given by 
\begin{align}
C (\alpha, \theta) & = \sum_n \left[ 
  (\cos(2\pi \alpha n + 2 \pi \theta) |n,L \rangle + \sin(2\pi \alpha n + 2 \pi \theta) |n,R \rangle)\langle n,L| \right. \notag \\
 & \ \ \ \ \ \left. + (\cos(2\pi \alpha n + 2 \pi \theta) |n,R \rangle - \sin(2\pi \alpha n + 2 \pi \theta) |n,L \rangle)\langle n,R| \right] \notag \\
& := \sum_n \kb{n} \otimes \hat{C}_n (\alpha, \theta),
\label{coin}
\end{align}
where 
$\alpha$ and $\theta$ are constant real numbers. 
Then this class of DTQW is called an inhomogeneous quantum walk (QW). 
This model is based on the idea of the Aubry-Andr\'e model~\cite{AA}, which provides a solvable example of metal-insulator 
transition in a one-dimensional incommensurate system. In this class of DTQW, we have obtained the weak limit theorem as follows. 
\begin{thm}[Shikano and Katsura~\cite{SK}]
	Fix $\theta = 0$. For any irrational $\alpha \in \RQ$ and any special rational $\alpha = \frac{P}{4Q} \in \mathbb{Q}$ 
	with relatively prime $P$ (odd integer) and $Q$, 
	the limit distribution of the inhomogeneous QW is given by 
	\begin{equation}
		\frac{X_t}{t^{\eta}} \Rightarrow I~~~~(t \to \infty),
		\label{limit}
	\end{equation}
	where $X_t$ is the random variable for the position at the $t$ step, ``$\Rightarrow$'' means the weak convergence, 
	and $\eta \ ( > 0)$ is an arbitrary positive parameter.
	Here, the limit distribution $I$ has the probability density function $f(x) = \delta (x)~~(x \in \R)$, 
	where $\delta ( \cdot )$ is the Dirac delta function. This is called a localization for the inhomogeneous QW.
	\label{inhom}
\end{thm}
However, in the case of the other rational $\alpha$, it has not yet been clarified whether the inhomogeneous QW is localized or not. This is still an open question. 
The situation becomes more complicated when we consider a nonzero $\theta$. 
As seen in Figure~\ref{open}, the reflection points for the quantum walker (see more details in Ref.~\cite[Lemma 1 and Figure 2]{SK})
are changed by the parameter $\theta$. 
In the rest of the paper, we will 
establish symmetry properties of the eigenvalue distribution of the one-step evolution operator ($U=WC$) at $\theta=0$. 

\begin{thm}
For the eigenvalues of the one-step evolution operator $WC$, 
the following properties hold:
	\begin{itemize}
		\item[(P1)] All the eigenvalues at $\alpha$ are identical to those at $1-\alpha$. 
		\item[(P2)] For every eigenvalue $\lambda$, there is an eigenvalue $\lambda^*$.
		\item[(P3)] For every eigenvalue $\lambda$, there is an eigenvalue $-\lambda$.
		\item[(P4)] All the eigenvalues are simple, i.e., nondegenerate.
		\item[(P5)] There are four eigenvalues $\lambda=\pm 1, \pm i$ for any $\alpha=\frac{P}{4Q} \in \mathbb{Q}$. 
		\item[(P6)] Every eigenvalue $\lambda$ at $\alpha=\frac{P}{4Q} \in \mathbb{Q}$ corresponds to an eigenvalue 
		$i \lambda$ at $\alpha + 1/2$.  
	\end{itemize}
\end{thm}
\begin{proof}
	The proofs of properties (P1) -- (P5) can be found 
	in Ref.~\cite{SK}. Here, we give a proof of (P6). According to Ref.~\cite[Theorem 3]{SK}, 
	the eigenvalues of $WC$ and $WC$ are identical. Therefore, we only study the eigenvalues of $CW$. First, we can express 
	the wavefunction at the $t$th step evolving from the state $\ket{0,{\tilde \phi}}$ by $CW$:
	\begin{equation}
		(CW)^t \ket{0,{\tilde \phi}} := \sum_{n \in \mathbb{Z}, \ \xi \in \{ L,R \} }\psiCW_t (n,\xi)\ket{n,\xi}.
	\end{equation}
	The one-step time evolution of the coefficients $\psiCW_t (n, \xi)$ is given by 
	\begin{equation}
		\left(\begin{array}{c} \psiCW_{t+1}(n; L) \\ 
		\psiCW_{t+1}(n; R) \end{array}\right)={\hat C}_n (\alpha, 0)
		\left(\begin{array}{c} \psiCW_{t}(n+1; L) \\
		\psiCW_t (n-1; R) \end{array}\right).
		\label{eq: CW_dynamics}
	\end{equation}
	Here, we define ${\vec{\psiCW}}_{t}$ by $\psiCW_t (n, \xi)$ and 
	a square matrix of order $4Q$, denoted as ${\sf CW}$, as 
	\begin{equation}
		{\vec{\psiCW}}_{t+1} = {\sf CW} {\vec{\psiCW}}_{t},
	\end{equation}
	see more details in Ref.~\cite{SK}.
	Let ${\vec{\psiCW}}=(\varphi(-Q; R), \varphi(-Q+1; L), \varphi(-Q+1; R), ..., \varphi(Q; L))^{{\bf T}}$ be the eigenvector of ${\sf CW}$ at 
	$\alpha$ with the eigenvalue $\lambda$ and ${\vec{\psitwoCW}}=({\tilde \varphi}(-Q; R), {\tilde \varphi}(-Q+1; L), {\tilde \varphi}(-Q+1; R), 
	..., {\tilde \varphi}(Q; L))^{{\bf T}}$ be at $\alpha + 1/2$ 
	with the eigenvalue $\tilde{\lambda}$.
	Then, according to Eq. (\ref{eq: CW_dynamics}), we obtain 
	\begin{align}
		\lambda \psiCW (-Q; R) &= (-1)^{\frac{P+1}{2}}\psiCW (-Q+1; L), \notag \\
		\lambda \left(\begin{array}{c} \psiCW (n; L) \\ 
		\psiCW (n; R) \end{array}\right) &= {\hat C}_n (\alpha, 0)
		\left(\begin{array}{c} \psiCW (n+1; L) \\ 
		\psiCW (n-1; R) \end{array}\right), ( n \in (-Q,Q)) \notag \\
		\lambda \psiCW (Q; L) &= (-1)^{\frac{P+1}{2}}\psiCW (Q-1; R) 
		\label{eq: CW_dynamics_2}
	\end{align}
	and 
	\begin{align}
		\tilde{\lambda} \psitwoCW (-Q; R) &= (-1)^{-Q} (-1)^{\frac{P+1}{2}}\psitwoCW (-Q+1; L), \notag \\
		\tilde{\lambda} \left(\begin{array}{c} \psitwoCW (n; L) \\ 
		\psitwoCW (n; R) \end{array}\right) &= (-1)^n {\hat C}_n (\alpha, 0)
		\left(\begin{array}{c} \psitwoCW (n+1; L) \\ 
		\psitwoCW (n-1; R) \end{array}\right), ( n \in (-Q,Q)) \notag \\
		\tilde{\lambda} \psitwoCW (Q; L) &= (-1)^{Q} (-1)^{\frac{P+1}{2}}\psitwoCW (Q-1; R),
		\label{eq: CW_dynamics_3}
	\end{align}
	where we have used the fact ${\hat C}_n (\alpha + 1/2) = (-1)^n {\hat C}_n (\alpha, 0)$. 
	Now we apply the following local unitary transformation to Eq. (\ref{eq: CW_dynamics_3}):
	\begin{equation}
		\psitwoCW (n; \xi) = \begin{cases} \psithreeCW (n; \xi) & {\rm when} \ n \ {\rm is \ even,} \\
		i \psithreeCW (n; \xi) & {\rm when} \ n \ {\rm is \ odd.}
		\end{cases}
		\label{trans}
	\end{equation}
	According to Eq. (\ref{eq: CW_dynamics_2}), $\vec{\psithreeCW}$ defined by Eq.~(\ref{trans}) can be taken as the eigenvector ${\sf CW}$ at $\alpha$ 
	with the eigenvalue $\tilde{\lambda} = i \lambda$.
\end{proof}
	
\begin{figure}[tb]
	\centering
	\includegraphics[width=10cm]{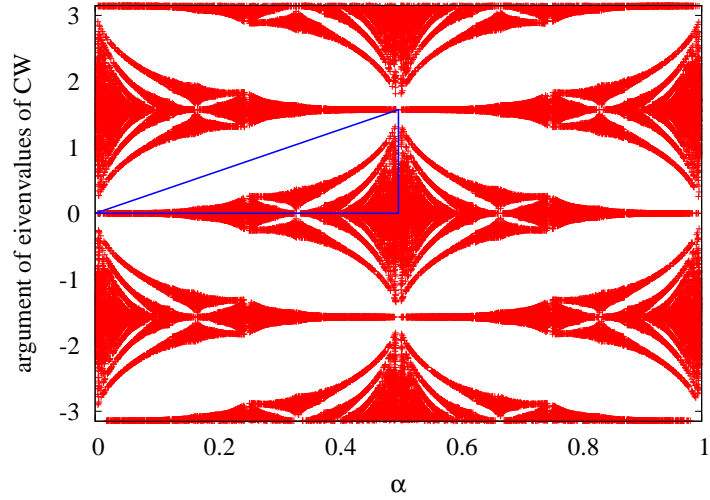}
	\caption{Eigenvalue distribution of the one-step operator for the inhomogeneous QW ($U$). 
	Arguments of the eigenvalues of $WC$ (vertical axis) are 
	plotted as a function of 
	the parameter $\alpha=\frac{P}{4Q}$ (horizontal axis) with $Q \le 60$. Here, $P$ (odd number) and $Q$ are relatively prime.}
	\label{dynamics}
\end{figure}

Figure \ref{dynamics} shows the numerically obtained spectrum of ${\sf CW}$ as a function of 
$\alpha$, which is quite similar to the Hofstadter butterfly~\cite{Hofstadter}.
By combining all the properties of (P1)-(P6), the smallest fundamental domain of this diagram is identified as the triangular region shown 
in Figure~\ref{dynamics}. Therefore, we have rigorously established all the symmetries 
in Figure~\ref{dynamics}. 
	
	One of the authors (Y.S.) thanks Shu Tanaka, Reinhard F. Werner and Volkher Scholz for useful discussions. 
	Y.S. is supported by JSPS Research Fellowships for Young Scientists (Grant No. 21008624). 
	H.K. is supported by the JSPS Postdoctoral Fellowships for Research Abroad and NSF Grant No. PHY05-51164.	

\end{document}